\documentclass[12pt]{article}
\usepackage{amsfonts,amsthm}
\usepackage{hyperref}

\oddsidemargin=\evensidemargin  
\newcommand{\la}{\lambda}
\newcommand{\om}{\omega}

\newcommand{\dl}{\delta}
\newcommand{\ba}{\begin{array}{l}}
\newcommand{\ea}{\end{array}}
\newtheorem{theorem}{Theorem}
\newtheorem{remark}{Remark}
\newtheorem{defin}{Definition}

\begin{document}

\title{Quantum multipole noise}
\author{Alexander Pechen\footnote{\href{mailto:pechen@mi.ras.ru}{pechen@mi.ras.ru}; \href{http://www.mathnet.ru/eng/person/17991}{www.mathnet.ru/eng/person/17991}}\\
\textit{Steklov Mathematical Institute of Russian Academy of Sciences,}\\ \textit{Gubkina 8, Moscow 119991, Russia}}
\date{}
\maketitle

\begin{abstract}Quantum multipole noise is defined as a
family of creation and annihilation operators with
commutation relations proportional to derivatives of delta
function of difference of the times, $[c^-_n(t),c^+_n(\tau)]\approx \delta^{(n)}(\tau-t)$. In this paper an explicit operator
representation of the quantum multipole noise is constructed
in a suitable pseudo-Hilbert space (i.e., in a Hilbert
space with indefinite metric). For making this representation, we introduce a class of Hilbert spaces obtained as completion of the Schwartz space in specific norms. Using this representation,
we obtain an asymptotic expansion as a series in quantum multipole
noise for multitime correlation
functions which describe the dynamics of open quantum systems
weakly interacting with a reservoir.
\end{abstract}

\section{Introduction}
We denote by $S(\mathbb R)$ the Schwartz space of complex
valued functions on $\mathbb R$ vanishing at infinity
faster than any polynomial~\cite{ref1}; by
$C^\infty(\mathbb R^d)$ the space of complex valued smooth
functions on $\mathbb R^d$; and by $C_0^\infty(\mathbb
R^d)$ the space of complex valued smooth functions with
compact support. Square brackets $[\cdot,\cdot]$ denote
commutator in a Lie algebra, i.e., $[A,B]=AB-BA$. The
anti-commutator is denoted as $\{A,B\}=AB+BA$.

\subsection{Classical white noise}
Classical white noise is a family of (in general,
complex-valued) random variables $w(t)$, $t\in\mathbb R$,
with mean $\mathbb E\{w(t)\}=0$ and the autocorrelation
$\mathbb E\{\bar w(t)w(\tau)\}=\dl(t-\tau)$. Here
$\dl(t-\tau)$ is the Dirac delta function and $\bar w$ is
the complex conjugate of $w$. Since the autocorrelation is
a distribution rather than a function, a proper definition
of the classical white noise should include choosing a
suitable space of test functions, e.g., the Schwartz space
$S(\mathbb R)$. Then, the classical white noise is defined
as a family of random variables $\{w(\phi)\,|\,\phi\in
S(\mathbb R)\}$ such that for any $\phi,\phi'\in S(\mathbb
R)$: $\mathbb E\{w(\phi)\}=0$ and $\mathbb E\{\bar
w(\phi)w(\phi')\}=\int dt \bar\phi(t)\phi'(t)$. Formally
one can set $w(\phi)=\int dt \phi(t) w(t)$.

\subsection{Quantum white noise}
Quantum white noise is defined in a similar way using a
suitable notion of a quantum probability space. We will
use the notion of a $*$-probability space.
\begin{defin}
A {\bf $*$-probability space} is a pair $({\cal A},\om)$,
where ${\cal A}$ is a unital $*$-algebra over $\mathbb C$
and  $\om:{\cal A}\to\mathbb C$ is a state, i.e., a
linear, normalized [$\om(1_{\cal A})=1$] and strictly
positive functional.
\end{defin}

Boson (resp., fermion) quantum white noise is defined as a
family of creation and annihilation operators (more
precisely, operator valued distributions) $a^+(t)$ and
$a^-(t)$ (where $t\in\mathbb R$) with the commutation
(resp., anti-commutation) relations proportional to
delta-function. Thus, boson quantum white noise satisfies
the relations
\begin{eqnarray}
~[a^-(t),a^+(\tau)]&=&\gamma_0 \dl(t-\tau)\label{eq3}\\
~[a^-(t),a^-(\tau)]&=&[a^+(t),a^+(\tau)]=0\label{eq4}
\end{eqnarray}
where $\gamma_0>0$. Respectively, fermi white noise the
relations
\begin{eqnarray*}
\{a^-(t),a^+(\tau)\}&=&\gamma_0 \dl(t-\tau)\\
\{a^-(t),a^-(\tau)\}&=&\{a^+(t),a^+(\tau)\}=0
\end{eqnarray*}
Again, since the commutator for both cases is a
distribution rather than a regular function, the proper
definition should include a suitable space of test
functions, e.g., the Schwartz space $S(\mathbb R)$. Then,
boson white noise is a family of operators
$\{a^\pm(\phi)\,|\, \phi\in S(\mathbb R)\}$ satisfying the
commutation relations~\cite{b,boloto}
\begin{eqnarray*}
~[a^-(\phi),a^+(\phi')]&=&(\phi,\phi')_{L^2(\mathbb R)}=\int dt\bar\phi(t)\phi'(t)\\
~[a^-(\phi),a^-(\phi')]&=&[a^+(\phi),a^+(\phi')]=0
\end{eqnarray*}
and similarly fermion quantum white noise satisfies the
anti-commutation relations for fermi case.

\subsection{Quantum multipole noise}
Quantum multipole noise is an operator valued distribution
with commutation relations proportional to derivatives of
delta function. The formal definition is as follows.
\begin{defin}
Quantum multipole boson noise is a (polynomial)
$*$-algebra generated by elements
$\{c^\pm_n(f)\,|\,n\in\mathbb N\bigcup\{0\},\, f\in
S(\mathbb R)\}$ with the commutation relations
\begin{eqnarray}
{}[c^-_m(f),c^+_n(h)]&=&\dl_{n,m}i^n\gamma_n\int\limits_{\mathbb
R}\overline{f^{(n)}(t)}h(t)dt\label{eq11}\\
{} [c^-_m(f),c_n^-(h)]&=&[c^+_m(f),c^+_n(h)]=0\label{eq2}
\end{eqnarray}
where $\delta_{n,m}$ is the Kronecker delta symbol,
$\gamma_n\ne 0$ are real numbers with $\gamma_0>0$,
$f^{(n)}$ and $\bar f$ denote $n$-th derivative and
complex conjugate of the function $f$. The involution $*$
is defined in the standard way as the extension of
$[c^-(f)]^*=c^+(f)$ to the whole algebra.
\end{defin}

We will use the formal notations $c^+(f)=\int
dtf(t)c^+(t)$ and $c^-(f)=\int dt\bar f(t)c^-(t)$, where
$c_n^\pm(t)$ are the operator valued distributions with
the commutation relations
\begin{eqnarray}\label{2}
~[c^-_n(t),c^+_m(\tau)]&=&\dl_{n,m}i^n\gamma_n\delta^{(n)}(\tau-t)\label{eq5}\\
~[c^-_n(t),c^-_m(\tau)]&=&[c^+_n(t),c^+_m(\tau)]=0\label{eq6}
\end{eqnarray}
and $\delta^{(n)}(\tau-t)=\partial^n_\tau\dl(\tau-t)$
denotes $n$-th derivative of delta function $\dl(\tau-t)$.

\begin{remark}
We call the operators $c_n^\pm(f)$ and the corresponding
operator valued distributions $c_n^\pm(t)$ as quantum
$2^n$-tuple noise. The reason is that the right hand side
(r.h.s.) of~(\ref{2}) for $m=n$ contains $n$-th derivative
of delta function and similarly, $n$-th derivative of
delta function determines charge density of a point
electric $2^n$-tuple moment.
\end{remark}

The case $n=0$ describes the standard quantum white noise,
i.e., the operator valued distribution with commutation
relations~\cite{alv}
\[
[c_0^-(t),c_0^+(\tau)]=\gamma_0\dl(\tau-t)
\]
Thus, $c^\pm_0(t)$ can be identified with the operators
$a^\pm(t)$ of boson quantum white noise. The simplest
nontrivial example of a quantum multipole noise
corresponds to $n=1$ and describes a quantum dipole noise
with commutation relations proportional to the first
derivative of delta function~\cite{pv,v}:
\[
[c_1^-(t),c_1^+(\tau)]=i\gamma_1\frac{\partial}{\partial\tau}\dl(\tau-t)
\]

Creation and annihilation operators of a quantum white
noise act in a standard, symmetric for the boson case and
antisymmetric for the fermion case, Fock space. In
contrast, as it will be shown below, the operators of
$2^n$-tuple noise for odd $n$ act in pseudo-Hilbert
spaces, i.e., in spaces with indefinite metric. An
operator representation of a quantum dipole noise in a
Fock space with indefinite metric was first constructed
in~\cite{pv}. In the next section we build an explicit
representation of the algebra of quantum multipole noise
by operators in an infinite tensor product of certain
Hilbert and pseudo-Hilbert Fock spaces.

\section{An operator representation of the quantum multipole noise}
In this section a representation of the quantum
$2^n$-tuple noise $c_n^\pm(f)$ is constructed for the case
$\gamma_n=1$ by creation and annihilation operators acting
in a symmetric Fock space with indefinite metric.

\begin{defin}
A pseudo-Hilbert space is a pair $({\cal H},\hat\eta)$,
where ${\cal H}$ is a (complex separable) Hilbert space
with positive defined inner product $(\cdot,\cdot)$ and
$\hat\eta: {\cal H}\to{\cal H}$ is a linear, bounded,
self-adjoint operator such that $\hat\eta^2=\mathbb I$,
where $\mathbb I$ is the identity operator in ${\cal H}$
(notice that such an operator must be unitary). The
operator $\hat\eta$ is called {\it metric operator}. The
indefinite inner product $\langle\cdot,\cdot\rangle$  for
any pair $f,h\in{\cal H}$ is defined as $\langle
f,h\rangle:=(f,\hat\eta h)$.
\end{defin}

An example of a space with indefinite metric is the
Minkowski space~\cite{ai}. For this example ${\cal H}=\mathbb R^4$
(real in this case) and $\hat\eta={\rm diag}(+1,-1,-1,-1)$
is the Minkowski metric. Clearly,
$\hat\eta^\dagger=\hat\eta$ and $\hat\eta^2=\mathbb I$.

\begin{remark}
The metric operator can be represented as a difference of
two projectors, $\hat\eta=\eta_+-\eta_-$, such that
$\eta_++\eta_-=\mathbb I$ [explicitly $\eta_\pm=(\mathbb
I\pm\eta)/2$]. This decomposition induces the
decomposition of the Hilbert space $\cal H$ in a direct
sum ${\cal H}={\cal H}_+\oplus{\cal H}_-$, where ${\cal
H}_\pm=\eta_\pm{\cal H}$ and for any $f\in{\cal H}_+$ and
$g\in{\cal H}_-$: $\langle f,f\rangle\ge 0$ and $\langle
g,g\rangle\le 0$.
\end{remark}

We will need the operator $F$ of the Fourier transform
\[
(Fh)(x)=\frac{1}{\sqrt{2\pi}}\int_{\mathbb R}
e^{itx}h(t)dt,\qquad h\in S(\mathbb R)
\]
and we denote $Fh=h_F$. Let ${\cal H}_n$ be the Hilbert
space constructed as the completion of the Schwartz space
$S({\mathbb R})$ with respect to the norm induced by the
following sesquilinear form:
\[
(f,h)_{{\cal H}_n} =\int_{\mathbb R}
|x|^n\overline{f_F(x)}h_F(x)dx,\qquad f,h\in S(\mathbb R)
\]
Thus, ${\cal H}_n=\overline{S(\mathbb
R)}^{\|\cdot\|_{{\cal H}_n}}$ where the norm
$\|\cdot\|_{{\cal H}_n}$ in $S(\mathbb R)$ is defined by
$\|f\|^2_{{\cal H}_n}=(f,f)_{{\cal H}_n}$. The same
notations $(\cdot,\cdot)_{{\cal H}_n}$ and
$\|\cdot\|_{{\cal H}_n}$ will be used in the sequel to
denote the inner product and the norm in the completed
${\cal H}_n$.

We define the metric operator $\eta_n$ in ${\cal H}_n$ as
a unique extension of the linear operator
$\eta=F^{-1}\circ{\rm sign}\circ F$ from the dense
subspace $S(\mathbb R)\subset{\cal H}_n$ onto ${\cal
H}_n$. Here $\rm sign$ is the multiplication operator by
the function ${\rm sign}(x)$: ${\rm sign}(x)=1$ if $x\ge
0$ and ${\rm sign}(x)=-1$ otherwise. Clearly, the operator
$\eta_n$ satisfies the properties of a metric operator:
$\eta_n=\eta_n^\dagger$ and $\eta_n^2=1$. The indefinite
inner product in ${\cal H}_n$ for odd $n$ has the form
$$
\langle f,h\rangle_{{\cal H}_n}:=(f,\eta_n h)_{{\cal
H}_n}= i^n\int\limits_{\mathbb
R}\overline{f^{(n)}(t)}h(t)dt
$$
For even $n$ we set $\langle f,h\rangle_{{\cal
H}_n}:=(f,h)_{{\cal H}_n}$ to be a positive defined inner
product. Now for any odd $n$ we have a pseudo-Hilbert
space $({\cal H}_n,\eta_n)$ and for any even $n$ a Hilbert
space ${\cal H}_n$.

For each $n$, define the symmetric Fock space over the
Hilbert space ${\cal H}_n$~\cite{boloto}
\[
{\cal F}_n:={\cal F}_{\rm sym}({\cal
H}_n)\equiv\bigoplus\limits_{k=0}^{\infty}{{\cal
H}_n}^{\otimes^k_{\rm sym}}
\]

We remind the following standard definition~\cite{boloto}.
\begin{defin} Let $\cal H$ be a Hilbert space.
A (symmetric) second quantization of a unitary operator
$U:{\cal H}\to{\cal H}$ is the unitary operator ${\cal
F}_{\rm sym}(U)$ in the symmetric Fock space
${\cal F}_{\rm sym}({\cal H})$ which acts in the $n$-particle subspace of ${\cal F}_{\rm sym}({\cal H})$ as $n$-th symmetric tensor power of $U$.
\end{defin}
With this definition, the indefinite metric in ${\cal
H}_n$ (for odd $n$) induces the indefinite metric
$\langle\cdot,\cdot\rangle_{{\cal F}_n}$ in the Fock space
${\cal F}_n$ by means of the metric operator
$\hat\eta_n:={\cal F}_{\rm sym}(\eta_n)$ which is defined as the
second quantization of the unitary $\eta_n$. Thus, for
each odd $n$ we have a pseudo-Fock space $({\cal
F}_n,\hat\eta_n)$ and for each even $n$ a Fock space
${\cal F}_n$.

Let $\Phi\in {\cal F}_n$ be a finite vector
$$
\Phi=(f_0,f_1(t_1),\dots,f_k(t_1,\dots,t_k),\dots)
$$
i.e., each $f_i(t_1,\dots,t_i)$ is a symmetric function
and $\exists N\in\mathbb N$ such that $f_k=0$ for all
$k\ge N$. In particular, the vacuum vector $\Phi^{\rm
vac}_n\in{\cal F}_n$ is defined as $\Phi^{\rm
vac}_n=(1,0,0,\dots)$. The set ${\cal D}_n\subset{\cal
F}_n$ of all finite vectors is a dense subset in ${\cal
F}_n$. Now define the action of the quantum $2^n$-tuple
noise operators $\{ c^\pm_n(f)\,|\,f\in S(\mathbb R)\}$ on
the set ${\cal D}_n$ by their action on $k$-particle
component of a vector $\phi\in{\cal D}_n$ as
\begin{eqnarray}
(c^+_n(f)f_k)_{k+1}(t_1,\dots,t_{k+1})&=&\frac{1}{\sqrt{k+1}}
\sum\limits_{i=1}^{k+1}f(t_i)f_k(t_1,\dots,\widehat{t}_i,\dots,t_{k+1})\label{c+}\\
(c^-_n(f)f_k)_{k-1}(t_1,\dots,t_{k-1})&=&i^n\sqrt{k}\int\overline{f^{(n)}(t)}
f_k(t,t_1,\dots,t_{k-1})dt\label{c}
\end{eqnarray}
The hat in~(\ref{c+}) means that the argument $t_i$ is
omitted.

\begin{theorem}\label{t1}
The operators $c^\pm_n(f)$ satisfy on the set of finite
vectors ${\cal D}_n\subset{\cal F}_n$ the commutation
relations
\begin{eqnarray*}
~[c^-_n(f),c^+_n(h)]&=&\langle f,h\rangle_{{\cal H}_n}\\
~[c^-_n(f),c^-_n(h)]&=&[c^+_n(f),c^+_n(h)]=0
\end{eqnarray*}
and the relation
$$
\langle c^-_n(f)\Phi,\Psi\rangle_{{\cal F}_n}
=\langle\Phi,c^+_n(f)\Psi\rangle_{{\cal F}_n},\qquad
\phi,\psi\in{\cal D}_n
$$
which means that the annihilation operator $c^-_n(f)$ is
(pseudo) adjoint to the creation operator $c^+_n(f)$.
\end{theorem}
\noindent {\bf Proof.} By direct calculations.

Let
\begin{equation}\label{eq1}
{\cal F}=\bigotimes\limits_{n=0}^\infty{\cal F}_n
\end{equation}
be the infinite tensor product of the Hilbert spaces
${\cal F}_n$ in the von Neumann sense~\cite{vN} with respect to the stabilizing sequence
$\{\Phi^{\rm vac}_n\}$ of vacuum vectors. Define in ${\cal
F}$ the metric operator $\hat\eta=\mathbb
I\otimes\hat\eta_1\otimes \mathbb
I\otimes\hat\eta_3\otimes\dots$ and identify the operators
$c^\pm_n(f)$ with the operators acting in ${\cal F}$ as
$c^\pm_n(f)$ in the $n$-th multiplier of the tensor
product in the r.h.s. of~(\ref{eq1}) and as the identity
in the other multipliers. The metric operator $\hat\eta$
introduces in $\cal F$ the structure of a pseudo-Hilbert
space $({\cal F},\hat\eta)$. The immediate consequence of
the Theorem~\ref{t1} is the following theorem.
\begin{theorem} The operators $\{c^\pm_n(f)\,|\, n\in\mathbb N\bigcup\{0\}, f\in S(\mathbb R)\}$
realize an operator representation of the quantum
multipole noise~(\ref{eq11},\ref{eq2}) in the
pseudo-Hilbert space $({\cal F},\hat\eta)$.
\end{theorem}
The generalization to arbitrary $\gamma_n$ is
straightforward.

\section{An asymptotic expansion for the multitime correlation functions}
Quantum multipole noises appear in the analysis of higher
order approximations to stochastic equations describing
dynamics of a quantum open system interacting with a
reservoir in the weak coupling limit (WCL). In this limit
the reduced dynamics of the system is described by
Markovian master equations~\cite{alv,ch,x}, and the total
dynamics of the system and the reservoir is governed by
white noise Schr\" odinger and quantum stochastic
differential equations.

Let ${\cal F}_{\rm sym}(L^2(\mathbb R^d))$ be the
symmetric Fock space over $L^2(\mathbb R^d)$ with the
vacuum vector $\Omega\in{\cal F}_{\rm sym}(L^2(\mathbb
R^d))$ and with the inner product $(\cdot,\cdot)$. The
boson creation and annihilation operators
$\{a^\pm(f)\,|\,f\in C_0^\infty(\mathbb R^d)\}$ are
defined in the usual way on the dense subspace of finite
vectors ${\cal D}\subset{\cal F}_{\rm sym}(L^2(\mathbb
R^d))$ and satisfy on this subspace the canonical
commutation relations
\[
[a^-(g),a^+(f)]=(g,f)_{L^2(\mathbb R^d)}
\]
and the relation
\[
(a^+(f)\Phi,\Psi)=(\Phi,a^-(f)\Psi),\qquad\Phi,\Psi\in{\cal
D}.
\]

Let $\om:\mathbb R^d\to\mathbb R$ be an infinitely
differentiable bounded from below function and
$\{S_t\}_{t\in\mathbb R}$ a one parameter unitary group in
$L^2(\mathbb R^d)$ acting on any $f\in L^2(\mathbb R^d)$
as $(S_tf)(p)=e^{it\om(p)}f(p)$. In the physical
applications, the function $\om$ is a dispersion law. For
example, for a two level system interacting with a
reservoir $\om(p)=p^2/2m-\om_0$, where $\om_0$ is the
difference between the the two energy levels of the system
and $p^2/2m$ is the kinetic energy of one reservoir
particle with mass $m>0$ and momentum $p\in\mathbb R^d$
($d=3$ in physical case).

Denote by $A_\la^\pm(t)=\la^{-1}a^\pm(S_{t/\la^2}g)$ the
rescaled free evolution of the creation and annihilation
operators. Let
\begin{equation}
\gamma_n=\frac{i^n}{n!}\int\limits_{-\infty}^{\infty}d\sigma\sigma^n\int\limits_{\mathbb
R^d}dke^{i\sigma\om(k)}|g(k)|^2.\label{eq7}
\end{equation}
Clearly, $\gamma_n$ for any natural $n$ is a real number
and $\gamma_0\ge 0$. Let $({\cal F},\hat\eta)$ be the
pseudo-Hilbert space with the indefinite inner product
$\langle\cdot,\cdot\rangle$ as defined in the previous
section and denote $\Phi_0=\Phi^{\rm
vac}_0\otimes\Phi^{\rm vac}_1\otimes\dots\otimes\Phi^{\rm
vac}_n\otimes\dots\in{\cal F}$.

\begin{theorem}\label{expA(t)}
Let $g\in C_0^\infty(\mathbb R^d)$ and $\om\in
C^\infty(\mathbb R^d)$ be such that ${\rm supp}\, g\cap
\{k_0\,|\,\nabla\, \om(k_0)=0\}=\emptyset$, where ${\rm
supp}\, g$ denotes the support of $g$. Then for any
natural numbers $n$ and $N$ the following equality holds
in the sense of distributions in $S'(\mathbb R^n)$ (in
variables $t_1,\dots, t_n$):
$$
(\Omega,A^{\epsilon_1}_\la(t_1)\dots
A^{\epsilon_n}_\la(t_n)\Omega)
=\Bigl\langle\Phi_0,\sum\limits_{i_1=0}^N\la^{i_1}
c^{\epsilon_1}_{i_1}(t_1)\dots\sum\limits_{i_n=0}^N\la^{i_n}
c^{\epsilon_n}_{i_n}(t_n)\Phi_0\Bigr\rangle+o(\la^N)
$$
where $\epsilon_i=\pm$ and $c^\pm_n(t)$ are the quantum
multipole noise operators (operator valued distributions)
satisfying the commutation relations~(\ref{2}) with
$\gamma_n$ defined by~(\ref{eq7}).
\end{theorem}

\noindent The proof follows from the Wick theorem~\cite{b}
and from the asymptotic expansion
\[
\frac{1}{\la^2}\int
dk|g(k)|^2e^{it\om(k)(t-\tau)/\la^2}\sim
\sum\limits_{n=0}^\infty(i\la)^n\gamma_n\dl^{(n)}(t-\tau).
\]
which is a consequence of the principle of locality~\cite{m,f}.
\begin{remark} The statement of the theorem can formally be interpreted as the asymptotic expansion
\begin{equation}\label{A}
 a^\pm(S_{t/\la^2}g)\sim\sum\limits_{n=0}^\infty\la^{n+1}c^\pm_n(t),\qquad\la\to 0.
\end{equation}
\end{remark}

\end{document}